\documentstyle [11pt,a4]{article}
\newcommand{\V}{{\bf V}}
\newcommand{\W}{{\bf W}}
\newcommand{\Na}{{\bf N}}
\newcommand{\si}{\sigma}
\newcommand{\pr}{\prime}

\newcommand{\be}{\begin{equation}}
\newcommand{\ee}{\end{equation}}
\newcommand{\bea}{\begin{eqnarray}}
\newcommand{\ena}{\end{eqnarray}}
\newcommand{\beas}{\begin{eqnarray*}}
\newcommand{\enas}{\end{eqnarray*}}

\newcommand{\hg}{U_q\hat{g}}
\newcommand{\g}{U_qg}

\newcommand{\D}{\Delta}
\newcommand{\Sp}{{\bf S}}
\newcommand{\ot}{\otimes}
\newcommand{\up}{\uparrow}
\newcommand{\dow}{\downarrow}

\newcommand{\nn}{\nonumber}
\newcommand{\la}{\lambda}

\newcommand{\hot}{\hat{\otimes}}

\newcommand{\op}{\oplus}
\renewcommand{\H}{{\cal H}}

\newcommand{\Pro}{{\cal Q}}
\newcommand{\Proj}{{\cal P}}
\newcommand{\ab}[1]{{\rm{\bf #1}}}
\renewcommand{\dim}{{\rm dim}}
\newcommand{\id}{{\rm id}}
\newcommand{\gr}[2]{ {( #1,#2)} }
\newcommand{\re}[1]{(\ref{#1})}

\begin{document}

\begin{center}
{ \large \bf
A New Family of Integrable Extended\\
Multi-band Hubbard Hamiltonians}\footnote{This work was supported in part by INTAS grant 94-840}

\vspace{36pt}

J.Ambjorn\footnote{e-mail:{\sl ambjorn@nbivms.nbi.dk}}

\vspace{6pt}
{\sl Niels Bohr Institute}

{\sl Blegdamsvej 17, Copenhagen, Denmark}

\vspace{24pt}

A.Avakyan\footnote{e-mail:{\sl avakyan@lx2.yerphi.am}},
            T.Hakobyan\footnote{e-mail:{\sl hakob@lx2.yerphi.am}}

\vspace{6pt}
{\sl Yerevan Physics Institute,}

{\sl Br.Alikhanian st.2, 375036, Yerevan, Armenia}

\vspace{24pt}

A.Sedrakyan\footnote{e-mail:{\sl sedrak@nbivms.nbi.dk}; permanent
address: Yerevan Physics Institute},

\vspace{12pt}

{\sl Niels Bohr Institute}

{\sl Blegdamsvej 17, Copenhagen,  Denmark}

\vspace{36pt}
January 1997



\vfill
{\bf Abstract}
\end{center}

We consider exactly solvable $1d$ multi-band fermionic Hamiltonians,
which have affine quantum group symmetry for all values of
the deformation. The simplest Hamiltonian is a multi-band $t-J$ model with
vanishing spin-spin interaction, which is the affinization of
an underlying $XXZ$ model. We also find a  multi-band
generalization of standard $t-J$ model Hamiltonian.

\vfill

\newpage

\section{Introduction}
In \cite{HS}  we developed the technique for construction of
spin chain Hamiltonians, which have the energy levels  of finite
$XXZ$ Heisenberg magnets with a degeneracy of levels due to an
affine quantum group symmetry. One can call this procedure
an affinization of $XXZ$ model.
The first example of this type model was constructed in
\cite{RA}, giving rise to  the Hubbard
Hamiltonian in the infinite repulsion limit.
In \cite{HS2} we have fermionized the simplest examples of this newly defined
family of spin chain models and have shown that it leads to
extensions of one-band Hubbard Hamiltonians. The $\eta$-pairing
mechanism introduced by Yang \cite{Y2,YZ} was found in one of
the examples, in addition to other extended exactly solvable Hubbard
models with super-conducting ground state \cite{St,Ov,BKS,MC,Schad}.

An interesting aspect of these models is that
the existence of an affine quantum group symmetry and the associated
degeneracy of levels might lead to
a new type of string theories.

In this article we extend the investigations to $t-J$ models initiated in
\cite{A,ZhR}. In sec.2 we review the construction of spin chain Hamiltonians
which possesses affine quantum group symmetry \cite{HS,HS2}.
In sec.3 we use this construction  to obtain a
multi-band $t-J$ model at $J=0$ which is the affinization of the $XXZ$
Heisenberg Hamiltonian. In sec.4 we show how a multi-band $t-J$ model is
obtained from the standard one-band $t-J$ model by the action of the
affine quantum group.
It is shown that some simple multi-band extension
of the $t-J$ model, where the spin-spin coupling term consists of the
interaction between the  total spins,
{\it i.e.\ the sum of the spins of all bands}, at nearest
 neighbor sites, contains an affine quantum group symmetry.

\setcounter{equation}{0}
\section{Quantum group invariant Hamiltonians for reducible
	representations}

Let $\V=\oplus_{i=1}^N \V_{\la_i}$ be a direct sum of finite dimensional
irreducible representations $\V_{\la_i}$ of quantum group $\g$ 
\cite{Dr86,J85, J86_1}.
We denote by $\V(x_1,\dots,x_N)$ the
representation with spectral parameters $x_i$
of the corresponding affine quantum group $\hg$ \cite{J86_1}:
\be
\label{Vdecompos}
 \V(x_1,\dots,x_N)=\bigoplus_{i=1}^M \Na_{\la_i}\hot \V_{\la_i}(x_i),
\ee
where all the $\V_{\la_i}(x_i)$ are $M$ nonequivalent irreps and
$\Na_{\la_i}\simeq \ab{C}^{N_i}$ have dimensions equal to
the multiplicity of $\V_{\la_i}(x_i)$
in $\V(x_1,\dots,x_N)$.
Note that $\sum_{i=1}^M N_i=N$.
The $\hat{~}$ over the tensor product signifies that $\hg$ does not act on
$\Na_{\la_i}\hot \V_{\la_i}(x_i)$ by means of comultiplication
$\D$ but instead acts as $\id\ot g$.

In \cite{HS2} the general matrix form of intertwining operator
\bea
\label{int}
 && H(x_1,\dots,x_N):\nn\\
&& \quad \V(x_1,\dots,x_N)\ot \V(x_1,\dots,x_N)\rightarrow
\V(x_1,\dots,x_N)\ot \V(x_1,\dots,x_N),
\\
 && \quad {[}H(x_1,\dots,x_N),\D(a){]}=0, \qquad \forall
 a\in\hg\nn
\ena
had been
written using the projection operators
\be
\label{proj}
X^a_b=|a\rangle\langle b|
\ee
Here the  vectors
$|a\rangle$ span the space $\V$. In accordance with the decomposition
\re{Vdecompos} we will
use the double index $a=\gr{n_i}{a_i}$, $i=1,\dots,M$
where the first
index  $n_i=1,\dots,N_i$ characterizes the multiplicity of
$\V_{\la_i}$ and the second one $a_i=1,\dots,\dim{\V_{\la_i}}$
is the vector index of $\V_{\la_i}$.
Then the intertwining operator \re{int}   is
\bea
\label{HactMat}
H(A,B)=
\sum_{i,j=1}^M
\left(
\sum_{n_i,n_j,m_i,m_j}
A_{ij}\mbox{}_{n_in_j}^{m_im_j}
\sum_{a_i,a_j}
X^\gr{n_i}{a_i}_\gr{m_i}{a_i} \ot X^\gr{n_j}{a_j}_\gr{m_j}{a_j}\right.
\\
\left.
+ \sum_{n_i,n_j,m_i,m_j}
B_{ij}\mbox{}_{n_in_j}^{m_im_j}
\sum_{a_i,a_j,a_i^\prime,a_j^\prime}
R_{ij}\mbox{}^{a_i a_j}_{a_i^\prime a_j^\prime }(x_i/x_j)
 X^\gr{n_j}{a_j^\prime}_\gr{m_i}{a_i}
\ot X^\gr{n_i}{a_i^\prime }_\gr{m_j}{a_j}
\right) ,
\nn
\ena
where the $R$-matrix
\[
R_{\V_{\la_i}\ot \V_{\la_j} }(x_i/x_j) |a_i\rangle\ot|a_j\rangle =
\sum_{a_i^\prime,a_j^\prime}
R_{ij}\mbox{}^{a_i a_j}_{a_i^\prime a_j^\prime }(x_i/x_j)
|a_i^\prime\rangle\ot|a_j^\prime\rangle .
\]
is the intertwining operator between two actions of affine quantum group
$\hg$ on  $\V_{\la_i}\ot \V_{\la_j}$, which are induced
correspondingly by comultiplication
$\D$ and opposite comultiplication $\bar{\D}$
\cite{Dr86,J85,J86_1}:
\[
R_{\V_{\la_i}\ot \V_{\la_j} }(x_i/x_j) \D(g)=
\bar{\D}(g) R_{\V_{\la_i}\ot \V_{\la_j} }(x_i/x_j).
\]
$A_{ij}$ and $B_{ij}$, $B_{ii}=0$ in \re{HactMat} are arbitrary matrices.
In general, $H(A,B)$ depends on deformation parameter $q$ of quantum
group, which is included in the $R$-matrix.
Note that  $R_{\V_{\la_i}\ot \V_{\la_j} }(x_i/x_j)$ does not depend
on $q$ and is identity
only if
$\la_i$ or $\la_j$ are trivial one-dimensional representations.
So, in the special
case, when the only nontrivial $R$-matrixes in \re{HactMat} are
between representations, one of which is trivial representation,
the expression of $H(A,B)$ doesn't depend on $q$. Then $H(A,B)$
commutes with the quantum group action for all values of
deformation parameter. In the following  we consider only this case.

Following \cite{RA,HS} we can from the operator $H$ construct
the following Hamiltonian acting on $\W=\V^{\ot L}$:
\footnote{Here and in the following we omit the dependence on $x_i$}
\be
\H=\sum_{i=1}^{L-1}H_{ii+1},
\label{genhamil}
\ee
where the indices $i$ and $i+1$ denote the sites where $H$
acts non-trivially. By the construction, $\H$ is quantum group invariant:
\bea
{[}\H,\D^{L-1}(g){]}=0, & \forall g\in \hg
\nn
\ena
Let us define the projection operators $\Pro^i$ on $\V$ for each class of
equivalent irreps $(\la_i,x_i)$, $i=1,\dots,M$
\bea
\Pro^iv_j=\delta_{ij}v_j, & \forall v_j\in V_{\la_j}(x_j) \nn\\
\sum_{i=1}^M \Pro^i=\id, & (\Pro^i)^2=\Pro^i\nn
\ena
Their action on $\W$ is given by
\[
\Pro^i=\sum_{l=1}^L \Pro^i_l
\]
It is easy to see that these projections commute with Hamiltonian ${\cal H}$
and quantum group $\hg$:
\bea
\label{com}
[\Pro^i,{\cal H}]=0, & [\Pro^i,\hg]=0
\ena
Denote by
$\W_{p_1\dots p_M}$
the subspace of $\W$ with values $p_i$
of $\Pro^i$ on it. Then we have the decomposition
\be
\label{wdecompos}
\W=\bigoplus_{\stackrel{p_1,\dots,p_M}{p_1+\dots+p_M=L}}\W_{p_1\dots p_M}
\ee

Let $\V^0$ be the linear space, spanned by the highest
weight vectors in $V$:
$$
\V^0:=\op_{i=1}^N v_{\la_i}^0,
$$
where  $v^0_{\la_i}\in
\V_{\la_i}$ is a highest weight vector. We also define  $\W^0:=\V^{0\ \ot L}$.
The space
$\W^0$ is $\H$-invariant.
For general $q$ the action of $\hg$ on $\W^0$ generate whole
space $\W$.
Indeed, the $\hg$-action on each state of type
$v^0_{\la_{i_1}}\ot\ldots\ot v^0_{\la_{i_L}}$ generates the space
$\V_{\la_{i_1}}\ot\ldots\ot \V_{\la_{i_L}}$, because the tensor product of
finite dimensional irreducible
representations of an affine quantum group is irreducible \cite{ChP}.

Consider now the subspace $\W^0_{p_1\dots p_M}=\W^0 \cap \W_{p_1\dots p_M}$.
According to (\ref{wdecompos}) we have the decomposition
\be
\label{w0decompos}
\W^0=\bigoplus_{\stackrel{p_1,\dots,p_M}{p_1+\dots+p_M=L}}\W^0_{p_1\dots p_M}.
\ee
Note that
$$
d_{p_1\dots p_M}:=\dim \W^0_{p_1\dots p_M}  =
\left(\begin{array}{c} L \\ p_1\ldots p_M
\end{array}\right)N_1^{p_1}\ldots N_M^{p_M} .
$$

Let us define by ${\cal H}_0$ the restriction of ${\cal H}$ on
$\W_0$: ${\cal H}_0:={\cal H}|_{\W_0}$.
It follows from
(\ref{com})  that  Hamiltonians ${\cal H}$ and ${\cal H}_0$ have
block diagonal form with respect to the decompositions (\ref{wdecompos}) and
(\ref{w0decompos}), respectively.
Every eigenvector
$w^0_{\alpha_{p_1\dots p_M}}\in
\W^0_{p_1\dots p_M}$ with energy value $E_{\alpha_{p_1\dots p_M}}$
gives rise to an irreducible $\hg$-multiplet
$\W_{\alpha_{p_1\dots p_M}}$
of dimension
\be
\label{lev}
\dim \W_{\alpha_{p_1\dots p_M}}=
                        \prod_{k=1}^M(\dim \V_{\la_k})^{p_k}
\ee
On $\W_{\alpha_{p_1\dots p_M}}$
 the Hamiltonian $\H$ is diagonal with eigenvalue
$E_{\alpha_{p_1\dots p_M}}$.
In particular, in the
case when all $\V_{\la_i}$ are equivalent, the degeneracy levels
are  the same for all $E_{\alpha_{p_1\dots p_M}}$ and are equal to
$(\dim \V_\la)^L$.

Now, let us assume we know  the energy spectrum
$E_{\alpha_{p_1\dots p_M}}$ for ${\cal H}_0$.
Then the statistical sum is given by
\be
\label{ZH0}
Z_{{\cal H}_0}(\beta)=\sum_{\stackrel{p_1,\dots,p_M}{p_1+\dots+p_M=L}}
\sum_{\alpha_{p_1\dots p_M}=1}^{d_{p_1\dots p_M}}
\exp(\beta E_{\alpha_{p_1\dots p_M}}),
\ee
and it follows that the statistical sum of ${\cal H}$ has the following form:
\be
\label{ZH}
Z_{{\cal H}}(\beta)=\sum_{\stackrel{p_1,\dots,p_M}{p_1+\dots+p_M=L}}
\prod_{k=1}^M(\dim \V_{\la_k})^{p_k}
\sum_{\alpha_{p_1\dots p_M}=1}^{d_{p_1\dots p_M}}
\exp(\beta E_{\alpha_{p_1\dots p_M}}).
\ee
So, if the underlying Hamiltonian ${\cal H}_0$ is integrable and
its eigenvectors and eigenvalues can be found, then
we know these  for ${\cal H}$ too.
Acting with the quantum group on all
eigenvectors of an energy level of $\H_0$ one obtains the whole
eigenspace of ${\cal H}$ for this level.


\section{Multi-band $t-J$ model with vanishing spin-spin coupling $J=0$}

Let us consider here the quantum group $U_q\widehat{sl}_2$.
We choose $\V=\V_0\oplus \V_j$
for decomposition \re{Vdecompos}, i.e.\ we take a direct sum of the trivial
spin-$0$ and the $2j+1$-dimensional spin-$j$ representation of
$U_qsl_2$. The
$R$-matrix in the second term in \re{HactMat} does not depend on $q$ and
spectral parameters $x_i$ and coincides with the identity, as it was
mentioned above. So, using \re{HactMat} and \re{genhamil}, we
obtain the following Hamiltonian
\bea
\label{RA0}
{\cal H}(t,V_1,V_2)=\sum_{i=1}^{L-1}
\left[
-t\sum_{p=1}^{2j+1}
(X_i\mbox{}^p_0 X_{i+1}\mbox{}^0_p+
X_{i+1}\mbox{}^p_0 X_{i}\mbox{}^0_p)
+ V_1 X_{i}\mbox{}^0_0 X_{i+1}\mbox{}^0_0 \right.\nn
\\
\left.+V_2\sum_{p,p^\prime=1}^{2j+1} X_i\mbox{}^p_p
X_{i+1}\mbox{}^{p^\prime}_{p^\prime}
\right]
\ena
The Hamiltonian  ${\cal H}=\sum_i H_{i i+1}$ was constructed
from the  operator $H=H_{i i+1}$, where $H$
can be written in the matrix form
\be
\label{R}
H= \left(
        \begin{array}{cccc}
                 V_1 & 0 & 0 & 0 \\
        0 & 0 & -t\cdot \id & 0\\
                0 & -t\cdot \id & 0 & 0\\
         0 & 0 & 0 & V_2\cdot\id
     \end{array}
\right).
\ee
The projection on the highest weight space coincides with the
constructing block of the $XXZ$ Hamiltonian in an external magnetic
field. This implies that the restriction of \re{RA0} to  the space $\W^0$ is
\bea
\label{XXZ}
\H_0(t,W_1,W_2)&=&\H_{XXZ}(t,\Delta,B)
\nn\\
&=&-\frac{t}{2}\sum_{i=1}^{L-1} \left(
\sigma^x_i \sigma^x_{i+1}+\sigma^y_i \sigma^y_{i+1} +
\Delta\sigma^z_i \sigma^z_{i+1} +\frac B2\sigma^z_i\right),
\ena
where
\be
\label{DeltaW}
\Delta=-\frac {V_1+V_2}{2t}, \qquad
B=\frac 2t(V_1-V_2)
\ee
For the special case $V_1+V_2=0$ $\H_0$ gives rise to the free
fermionic (or equivalently $XY$) Hamiltonian ($\Delta=0$).

The projection operators $X^a_b$ are expressed through the
fermionic creation-an\-ni\-hi\-la\-tion operators as follows
\bea
\label{proj2}
 X_i\mbox{}^p_0= \Proj c_{i,p}^+, \qquad &
 X_i\mbox{}_p^0=c_{i,p}\Proj, \nn\\
X_i\mbox{}^p_p=n_{i,p}\Proj=\Proj n_{i,p}, &
X_i\mbox{}^0_0= (1-n_{i})\Proj=\Proj (1-n_i)
\ena
Here we introduced the projection operator which forbids double occupation
on all sites
$$
 \Proj=\prod_{i=1}^L { \Proj}_i, \qquad 
\Proj_i=\prod_{p\ne p^\prime}
(1-n_{i,p}n_{i,p^\prime})
$$
and the total particle number $n_i=\sum_pn_{i,p}$ at site $i$.

After the substitution of the fermionic representation \re{proj2}
into \re{RA0} we obtain
\bea
\label{RA1}
{\cal H}(t,V_1,V_2) = {\cal P}\sum^{L-1}_{i=1} \left[ -t
\sum_{p=1}^{2j+1}
 ( c^{+}_{i,p}
c_{i+1,p} + c^{+}_{i+1,p}c_{i,p}  )\right.
+\nn\\
\left. V n_in_{i+1}-V_1(n_i+n_{i+1})+V_1\right] {\cal P},
\ena
where $V=V_1+V_2$.
The chemical potential term $-V_1\sum_{i=1}^{L-1}(n_i+n_{i+1})$
commutes with ${\cal H}$ and can be omitted. So, up to
unessential boundary and constant terms \re{RA1} is a multicomponent
$t-J$ model with vanishing spin-spin coupling ($J=0$)
\bea
\label{RA}
{\cal H}(t,V) = \sum^{L-1}_{i=1} \left[ -t
\sum_{p=1}^{2j+1}
 ( c^{+}_{i,p}c_{i+1,p} + c^{+}_{i+1,p}c_{i,p}  ) + Vn_in_{i+1} \right]
+\sum_{i=1}^L \sum_{\stackrel{p\ne p^\prime,}{p,p^\prime=1}}^{2j+1} 
U_{p,p^\prime}
n_{i,p}n_{i,p^\prime},
\ena
where the infinite Hubbard interaction amplitude
$U_{p,p^\prime}=+\infty$
 between  $p$ and $p^\prime$ bands excludes sites with double and
 more occupations.
 It follows from the above considerations that  this model has
 energy levels which  coincide with the levels of $XXZ$ Heisenberg
 model, but that the degeneracy of the levels is different.

For vanishing density-density interaction $V=0$ the Hamiltonian
\re{RA} describes the infinite repulsion limit of the multi-band
Hubbard model. Thus, according to \re{DeltaW} $\Delta=0$
and it has the energy levels of free fermionic model.

\section{Multi-band extension of $t-J$ model, which has affine
quantum group symmetry}

In this section we consider Hamiltonians which have the
same energy levels as $t-J$ model but have affine quantum group
symmetry. Because each site in ordinary
$t-J$ model has three states, one should for this purpose take
direct sum of three spaces. Let
\be
\label{t-J-de}
\V=\V_0\oplus \V_{j}\oplus \V_{j}
\ee
Recall the $t-J$ model is given by
\be
\label{t-J}
\H_{t-J}(t,J,V)=
{\cal P}\sum^{L-1}_{i=1} \left[ -t
\sum_{\sigma=\pm\frac12}
 ( c^{+}_{i,\sigma}
c_{i+1,\sigma} + c^{+}_{i+1,\sigma}c_{i,\sigma}  ) +J \Sp_i\Sp_{i+1}
+V n_in_{i+1}\right] {\cal P},
\ee
where $c^+_{\sigma}, \ c_\sigma$ are creation-annihilation operators of 
spin-$\frac12$ fermion, $\Sp=\sum_{\sigma,\sigma^\prime}
c^+_\sigma {\bf\sigma}_{\sigma\sigma^\prime}c_{\sigma^\prime}$
is the fermionic spin operator and ${\cal
P}=\prod_{i=1}^L(1-n_{i,\up}n_{i,\dow})$ forbids  double
occupation of sites.

We  rewrite it in terms of Hubbard operators $X^a_b$, where $a,b=0,\pm\frac12$:
\bea
\label{t-J-X}
{\cal H}(t,J,V)=\sum_{i=1}^{L-1}
\left[
\sum_{\sigma=\pm\frac12}\left(-t
(X_i\mbox{}^\sigma_0 X_{i+1}\mbox{}^0_\sigma+
X_{i+1}\mbox{}^\sigma_0 X_{i}\mbox{}^0_\sigma)
+ \frac 12 J\cdot X_{i}\mbox{}^\sigma_{-\sigma}
X_{i+1}\mbox{}^{-\sigma}_{\sigma} \right)\right.\nn
\\
\left.
+\sum_{\sigma,\sigma^\prime=\pm\frac12}
\left({\sigma\sigma^\prime} J+V\right)
X_i\mbox{}^\sigma_\sigma
X_{i+1}\mbox{}^{\sigma^\prime}_{\si^\pr}
\right]
\ena
Let us now look at the  general expression \re{HactMat} of intertwining
operators $H_{ij}$ acting on the space \re{t-J-de}.
For convenience we make index change in the following way.
The two spin-$j$ representations we use are denoted by $\sigma=\pm\frac12$.
The intrinsic index in each $V_j^{(\si)}$
is denoted  by $k$, $k=1,\dots,2j+1$.
So, instead of $(n_i,a_i)$ in \re{HactMat}
we have $(\si,k)$, if $i$ corresponds to  spin-$j$ multiplet.
Because the
spin-$0$ singlet is one dimensional and single, we just use for it the
index $0$.  The non-equivalent irreps in \re{t-J-de} are $V_j^{(\si)}$
and $V_0$ and, as mentioned above, the $R$-matrix for two such representations
 is the identity.  After performing the first sum in
\re{HactMat} over non-equivalent multiplets
we obtain
\bea
\label{HactMat-t-J}
H(A,a,b_1,b_2)=
\sum_{\si_1,\si_2,\si_1^\pr,\si_2^\pr}
\left(
A_{}\mbox{}_{\si_1\si_1^\pr}^{\si_2\si_2^\pr}
\sum_{k,k^\pr}
X^\gr{\si_1}{k}_\gr{\si_2}{k} \ot
X^\gr{\si_1^\pr}{k^\pr}_\gr{\si_2^\pr}{k^\pr}
\right)+ a\cdot X^0_0\ot X^0_0
\nn   \\
+\sum_{k,\sigma}
 \left(b_1 \cdot
X^\gr{\si}{k}_{0} \ot X^{0 }_\gr{\si}{k}
+b_2 \cdot X^0_\gr{\si}{k} \ot X^\gr{\si}{k}_0
\right)
\ena
To implement the  restriction $H_0(A,a,b_1,b_2)$ of this operator
on the highest weight space one just should eliminate the sum over $k,k^\prime$
 and put $k=k^\prime=0$. Comparing \re{HactMat-t-J} and \re{t-J-X} it follows that
the expressions coincide if one chooses
\[
a=0 \qquad b_1=b_2=-t \qquad
A_{\si-\si}^{-\si\si}=J/2
\qquad
A_{\si\si}^{\si^\pr\si^\pr}=(\si\si^\pr)\cdot J+V
\]
and choose the other values of $A_{\si_1\si_2}^{\si_1^\pr\si_2^\pr}$
equal zero.

So, the Hamiltonian $\H(A,a,b_1,b_2)$ corresponding to
\re{HactMat-t-J} with these values of parameters
gives rise to a $t-J$ model \re{t-J}  on the highest weight space.
According to the previous considerations it will have the same energy
levels as $t-J$ model, but  with different degeneracy. Recall that
for $J=2t$ the $t-J$ model is "supersymmetric" and integrable.

We express the Hubbard operators in terms of multi-band fermionic
creation\--an\-ni\-hi\-la\-tion operators
as follows
\bea
\label{proj3}
 X_i\mbox{}^\gr{\sigma}{k}_0=\Proj c_{i,\sigma}^{k+}, \qquad &
 X_i\mbox{}_\gr{\sigma}{k}^0=c_{i,\sigma}^k\Proj ,  \nn\\
\\
 X_i\mbox{}^\gr{\sigma}{k}_\gr{-\sigma}{k}
=c_{i,\sigma}^{k+}c_{i,-\sigma}^k\Proj=
\Proj c_{i,\sigma}^{k+}c_{i,-\sigma}^k,
& X_i\mbox{}^\gr{\sigma}{k}_\gr{\sigma}{k}
=n_{i,\sigma}^k\Proj=\Proj n_{i,\sigma}^k
\nn 
\ena
Here as before we used the projection operator, which forbids double occupation
on all sites
$$
 \Proj=\prod_{i=1}^L { \Proj}_i, \qquad 
\Proj_i=\prod_{\gr{\sigma}{k}\ne \gr{\sigma^\prime}{k^\prime}}
(1-n_{i,\sigma}^k n_{i,\sigma^\prime}^{k^\prime})
$$
Now, we can write down the
Hamiltonian \re{genhamil} in terms of
multi-band fermions, substituting \re{proj3} into \re{HactMat-t-J}.
We obtain in this way the multi-band
generalization of \re{t-J}
\be
\label{Mu-t-J}
\H(t,J,V)=
{\cal P}\sum^{L-1}_{i=1} \left[ -t
\sum_{k=1}^{2j+1}\sum_{\sigma=\pm\frac12}
 ( c^{k+}_{i,\sigma}
c^k_{i+1,\sigma} + c^{k+}_{i+1,\sigma}c^k_{i,\sigma}  ) +J
\Sp_i\Sp_{i+1}
+V n_in_{i+1}\right] {\cal P},
\ee
Here $k$ is the band index, and $\Sp=\sum_k\Sp^k$, $n=\sum_k
n^k$ are total spin and total particle number operators.


\end{document}